# Ferroelasticity in Two-Dimensional Hybrid Ruddlesden–Popper Perovskites Mediated by Cross-Plane Intermolecular Coupling and Metastable Funnel-Like Phases

Devesh R. Kripalani,[1]† Qiye Guan,[2]† Hejin Yan,[2]† Yongqing Cai,[2]* and Kun Zhou,[1,3]*

[1] *School of Mechanical and Aerospace Engineering, Nanyang Technological University, 50 Nanyang Avenue, Singapore 639798, Singapore.*

[2] *Joint Key Laboratory of the Ministry of Education, Institute of Applied Physics and Materials Engineering, University of Macau, Avenida da Universidade, Taipa, Macau 999078, P. R. China.*

[3] *Environmental Process Modelling Centre, Nanyang Environment and Water Research Institute, Nanyang Technological University, 1 Cleantech Loop, Singapore 637141, Singapore.*

† These authors contributed equally.

\* Corresponding authors. E-mail addresses: yongqingcai@um.edu.mo (Yongqing Cai), and kzhou@ntu.edu.sg (Kun Zhou).

**Abstract**

Ferroelasticity describes a phenomenon in which a material exhibits two or more equally stable orientation variants and can be switched from one form to another under an applied stress. Recent works have demonstrated that two-dimensional layered organic–inorganic hybrid Ruddlesden–Popper perovskites can serve as ideal platforms for realizing ferroelasticity, however, the ferroelastic (FE) behavior of structures with a single octahedra layer such as $(BA)_2PbI_4$ (BA = $CH_3(CH_2)_3NH_3^+$) has remained elusive. Herein, by using a combined first-principles and metadynamics approach, the FE behavior of $(BA)_2PbI_4$ under mechanical and thermal stresses is uncovered. FE switching is mediated by cross-plane intermolecular coupling, which could occur through multiple rotational modes, rendering the formation of FE domains and several metastable paraelastic (PE) phases. Such metastable phases are akin to wrinkled structures in other layered materials and can act as a "funnel" of hole carriers. Thermal excitation tends to flatten the kinetic barriers of the transition pathways between orientation variants, suggesting an enhanced concentration of metastable PE states at high temperatures, while halogen mixing with Br raises these barriers and conversely lowers the concentration of PE states. These findings reveal the rich structural diversity of $(BA)_2PbI_4$ domains, which can play a vital role in enhancing the optoelectronic properties of the perovskite and raise exciting prospects for mechanical switching, shape memory, and information processing.

**Keywords**

**Introduction**

Analogous to its sister ferroic properties, ferroelectricity, ferromagnetism, and ferrotoroidicity, ferroelasticity refers to the existence of two or more equally stable orientation variants of a crystal in the absence of mechanical stress which can be switched from one form to another without diffusion under the influence of an applied stress (*1–3*). These orientation variants correspond to different states of spontaneous strain in the crystal's ferroelastic (FE) phase and typically form through a symmetry-breaking structural phase transition from a parent paraelastic (PE) phase. It has been shown that FE domains of different orientations can coexist to form complex domain patterns and twin boundaries with intriguing functionalities (*1*, *4*). Moreover, under appropriate mechanical or thermal stimuli, the switching between different orientation variants can be achieved. This gives rise to a hysteric stress–strain behavior (*1*, *2*, *5*) as well as allows for the dynamic tuning of material properties (*6–10*).

Two-dimensional ferroelastic (2D-FE) materials hold great promise for next-generation nanoscale devices, with potential applications in mechanical switching, shape memory, and information processing (*6*, *8–11*). To date, a wide variety of 2D-FE materials have been identified, including transition metal dichalcogenides (TMDs) (*11*, *12*), group-IV monochalcogenides (*9*, *10*), phosphorene (*10*), α-SnO (*13*), and even boron-containing compounds such as borophane (*14*) and $BP_5$ (*15*). The study of such systems has not only bestowed scientists with a deeper understanding of their ferroelastic switching mechanism and orientation-dependent properties (electronic, mechanical, thermal, *etc*.), but has also disclosed their multiferroic behavior, which, in essence, involves the coupling between ferroelasticity and the other ferroic orders (ferroelectricity, ferromagnetism, *etc*.) of the system. Recently, experimental evidence of ferroelasticity in two-dimensional (2D) layered organic–inorganic hybrid Ruddlesden–Popper perovskites (HRPPs) has surfaced (*16*, *17*). Their structures resemble that of a quantum well configuration, consisting of a semiconducting inorganic octahedra framework sandwiched between insulating organic buffer layers, as given by the general formula R$_2$A$_{n-1}$M$_n$X$_{3n+1}$, where R refers to an organic cation spacer molecule, A is an interior monovalent cation, M is a divalent metal, and X is a halide, while *n* indicates the thickness of the octahedra network. A compelling understanding of the ferroelastic behavior of HRPPs will therefore be a key step towards realizing their utility for advanced, multifunctional applications.

By employing scanning tunneling microscopy, Telychko *et al*. successfully imaged the FE domains and twin boundary composition of the (BA)$_2$MA$_3$Pb$_4$I$_{13}$ (*n* = 4; BA = CH$_3$(CH$_2$)$_3$NH$_3^+$, MA = CH$_3$NH$_3^+$) perovskite and showed that these domains stem from the synergetic alignment of the interior polar MA chains and distortion of the Pb–I lattice (*16*). Strain-induced ferroelastic domain wall motion had also been observed in *n* = 2 and *n* = 3 homologues by Xiao *et al*. using a polarized optical microscope (*17*). Interestingly, however, there was no evidence of ferroelasticity in perovskites with a single octahedra layer (*n* = 1). Although this was ascribed to the lack of interior-site cations, perturbations of the organic spacer molecules could in principle also initiate the switching between different phases (*18–21*). Particularly, when the thickness of the perovskite is reduced into the monolayer, the spacer molecules will no longer be restrained by interlayer van der Waals (vdW) forces and they will be free to reorganize themselves under external stress (*22*). The existence of ferroelasticity in HRPPs with an *n* = 1 architecture and the mechanism by which it can manifest thus remain open questions of interest in the search for ultrathin 2D-FE perovskites.

In this work, based on density functional theory (DFT) simulations, we uncover ferroelasticity in (BA)$_2$PbI$_4$, an archetypical member of the *n* = 1 HRPP family. The switching between different orientation variants is regulated by the coupling between organic molecules of the top and bottom surfaces. There are multiple alternative pathways by which ferroelastic switching could take place depending on the rotational mode of the BA molecules. Notably, it is shown that during this process, the cooperative reordering of BA molecules can lead to the formation of metastable paraelastic phases. The activation free energy of variant switching can be manipulated by halogen alloying and the entropy contribution associated with the rotational freedom of the organic BA molecules. Thus, the energy landscape enabling ferroelastic response in such hybrid systems may be renormalized under different temperatures and through compositional engineering. Our findings reveal the rich structural diversity of $(BA)_2PbI_4$ domains, which can play a vital role in optoelectronic conversion and raise fresh perspectives for reconfigurable devices.

**Results and Discussion**

**Ferroelastic Order in Monolayer $(BA)_2PbI_4$**

Herein, we consider $(BA)_2PbI_4$ in its fundamental monolayer as a representative system to investigate the possibility of ferroelasticity in $n = 1$ HRPPs. It has been confirmed from experiments that $(BA)_2PbI_4$ crystallizes in the centrosymmetric orthorhombic space group *Pbca* with bulk lattice constants $a$ = 8.68–8.69 Å, $b$ = 8.86–8.88 Å, and $c$ = 27.57–27.63 Å (*22–24*). Moreover, its 2D layered structure allows for the facile isolation of single- and few-layer nanosheets *via* mechanical exfoliation (*22*, *25*). In the monolayer limit, $(BA)_2PbI_4$ adopts the same orthorhombic crystal structure, albeit with reduced lattice constants in the absence of interlayer vdW interactions. The lattice constants of monolayer $(BA)_2PbI_4$, as calculated in this work, are $a$ = 8.59–8.68 Å and $b$ = 8.67–8.73 Å, in good agreement with previously reported values ($a$ = 8.67–8.68 Å, $b$ = 8.80–8.87 Å) (*22*, *26*). Although slight imaginary frequencies (within −10 cm$^{-1}$) emerge in the calculated phonon spectrum of the monolayer at 0 K (Fig. S1), these modes can be stabilized through both in-plane and out-of-plane octahedral tilting along with the rotation of organic cations at higher temperatures. The dynamical stability of monolayer (BA)$_2$PbI$_4$ at room temperature can be verified from our molecular dynamics simulation results at 200 K and 300 K (which will be discussed in detail later), where it is shown that the monolayer flake does indeed remain intact throughout 10 ps-long simulations.

In view of the crystal symmetry of the monolayer, two energetically degenerate orientation variants I and II can be defined, as illustrated in Figs. 1(a) and 1(b), respectively. Note that the lattice constant along $x$ ($y$) is denoted by $a$ ($b$), and an accompanying subscript is used to specify which variant the parameter belongs to. The superscript "0" is used hereafter to refer to the groundstate condition. It can be seen that the variants I and II are identical from a crystallographic standpoint, relating to one another by an in-plane ($xy$-plane) rotation of 90°. The structural anisotropy of each variant originates from the different ways in which the BA molecules are arranged across the surface along either orthogonal direction (denoted by gray-shaded boxes). As shown, in variant I (II), the ammonium heads of adjacent BA molecules on both the top and bottom surfaces are aligned in a head-to-tail manner along $x$ ($y$), but in a head-to-head manner along $y$ ($x$). Theory dictates that $a_I = b_{II}$ and $b_I = a_{II}$ in the groundstate configurations (I$^0$ and II$^0$), however, slight numerical deviations do arise, and inevitably so, during geometry optimization because of the high rotational freedom of BA molecules. The lattice constants of I$^0$ and II$^0$ are evaluated to be $a_I^0 = 8.59$ Å, $b_I^0 = 8.73$ Å and $a_{II}^0 = 8.67$ Å, $b_{II}^0 = 8.68$ Å, respectively, while the energy difference between the two FE groundstates is calculated to be less than 0.3 meV/atom. The errors in optimization are small and can be neglected for the purposes of our work. This is corroborated by the fact that they have no discernible impact on the electronic band structure and density of states distribution of the monolayer (Fig. S2).

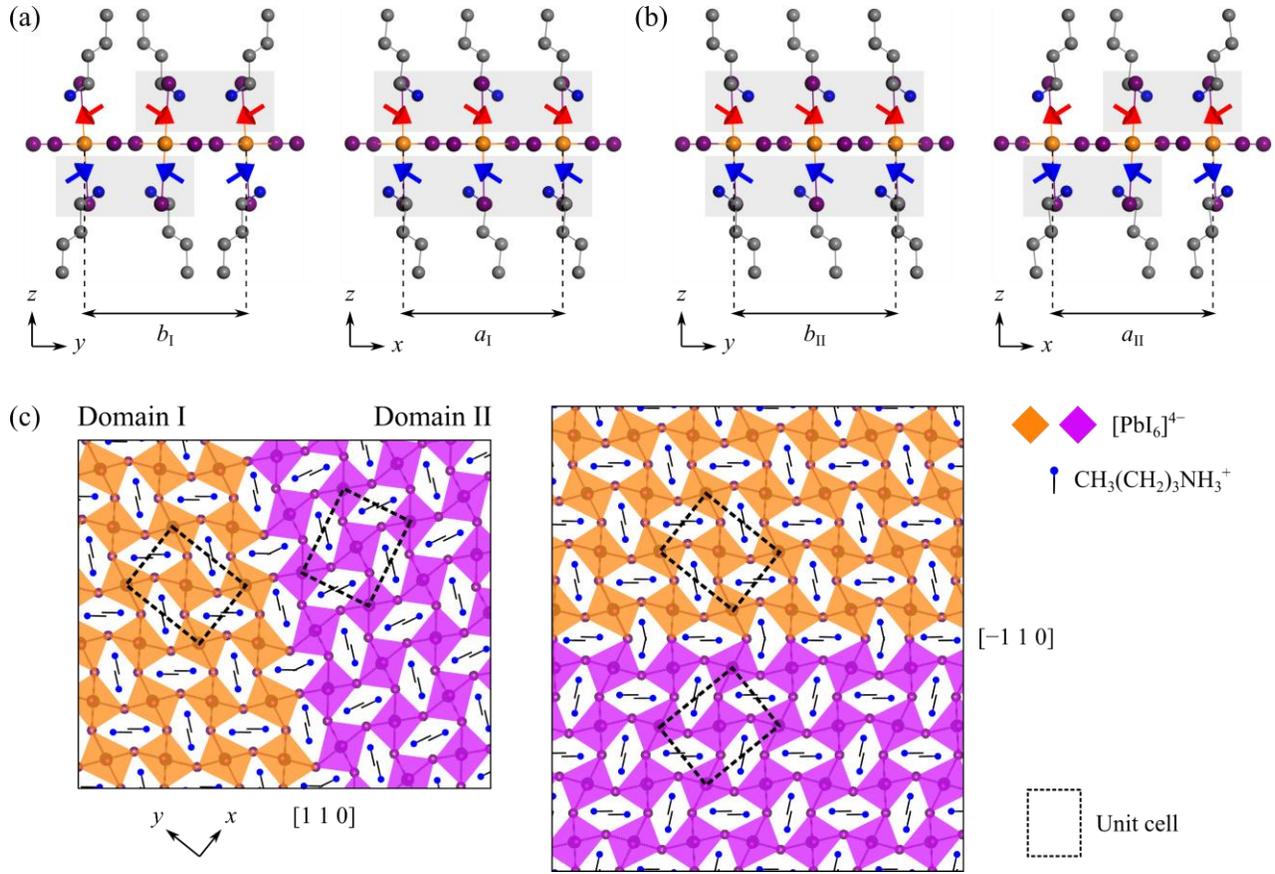

**Fig. 1.** Crystal structures (side view) of the ferroelastic orientation variants (a) I, and (b) II of monolayer $(BA)_2PbI_4$, where C, N, Pb, and I atoms are indicated in gray, blue, orange, and purple, respectively. Note that H atoms are omitted for clarity. Red (blue) arrows denote the C–N dipole orientations of BA molecules located on the top (bottom) surface of the monolayer. (c) Schematic illustration of the ferroelastic domain boundaries (top view).

A direct consequence of ferroelasticity in 2D-FE materials is the coexistence of domains of different orientations and the formation of quasi-one-dimensional twin boundaries between them. As shown schematically in Fig. 1(c), monolayer $(BA)_2PbI_4$ can harbor both variants I and II simultaneously as separate domains to form coherent twin boundaries with mirror planes aligned parallel to the [1 1 0] and [−1 1 0] directions (defined relative to variant I). Accordingly, the domain pattern would feature adjacent twin boundaries intersecting at near-90° angles. Under suitable mechanical or thermal stimuli, one of the variants could be stabilized at the expense of the other, resulting in twin boundary migration and the growth of a dominant orientation state. Ferroelastic switching between variants I and II essentially involves swapping the modes of BA alignment along the principal axes (*i.e.*, switching between head-to-tail and head-to-head arrangements). Specifically, such a transformation can be achieved through an in-plane rotation of a pair of BA molecules per unit cell by about 180°, as shown in Fig. 2(a). Highly coordinated molecular rotations across the surface could thus initiate a reversible transition from variant I to II or *vice versa*, thereby endowing the 2D perovskite with dual-state ferroelasticity.

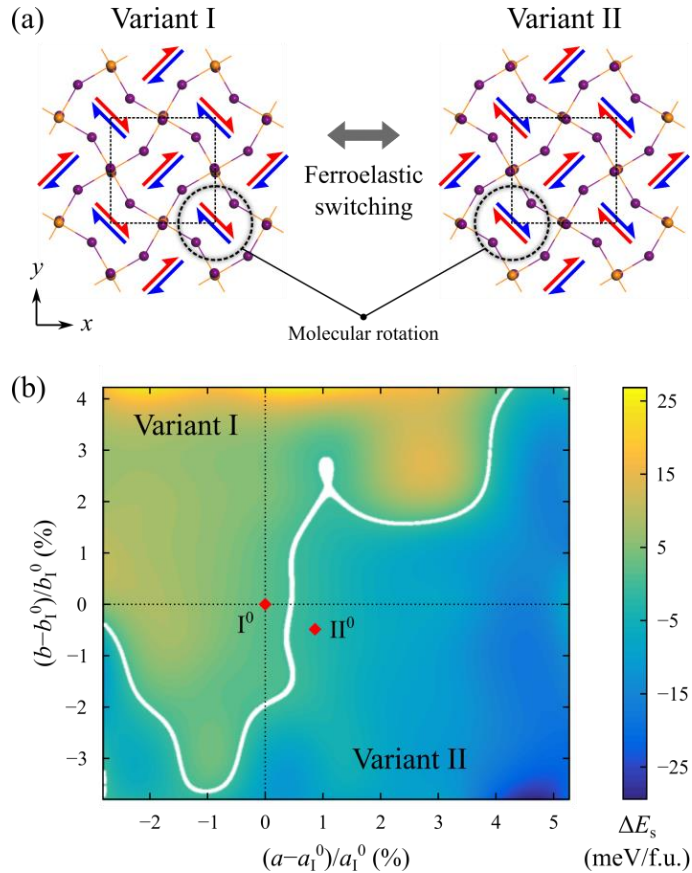

**Fig. 2.** (a) Schematic diagram of dual-state ferroelasticity in monolayer $(BA)_2PbI_4$. (b) Map of the strain energy difference $\Delta E_s$ between variants I and II as a function of lattice strain. The ferroelastic groundstates $I^0$ and $II^0$ are marked by red dots, while the intersection contour ($\Delta E_s \approx 0$ eV) between the two variants is indicated in white.

To investigate the energy landscape of variants I and II under applied strain and compare their relative stabilities across a broad range of lattice constants, we examine the amount of strain energy $E_s$ stored in each variant as a function of its lattice parameters (generalized here as $a$ and $b$), as given by

$$E_s(a,b) = E(a,b) - E^0, \quad (1)$$

where $E^0$ and $E$ are the total energies of the system in its groundstate and strained configuration, respectively. The orientation variant that minimizes $E_s$ is ideally the one that will be energetically favored. Note that, in our analysis, the amount of strain imposed onto the monolayer is quantified with respect to the groundstate lattice constants of variant I (*i.e.*, $a_I^0$ and $b_I^0$). The total energy $E$ of each variant is first computed on a 9 × 9 grid in ($a$, $b$) space with a step size of 1%, giving a total number of 81 data points sampled around the groundstate. Then, by approximating the intermediate values of $E(a, b)$ using 2D spline interpolation, we are able to derive smooth strain energy surfaces for each of the variants in the investigated ($a$, $b$) space. Finally, an intersection contour can be constructed based on the strain energy difference $\Delta E_s$ between the two variants (Fig. 2(b)), as given by $\Delta E_s = E_s^{II} - E_s^{I}$, where $E_s^{I}$ and $E_s^{II}$ refer to the strain energies per formula unit (f.u.) of variants I and II, respectively. The FE groundstates $I^0$ and $II^0$ are marked by red dots in Fig. 2(b), while the intersection contour ($\Delta E_s \approx 0$ eV) between the two variants is indicated in white. Positive (negative) values of $\Delta E_s$ signify that the structural configuration of variant I (II) is capable of minimizing the strain energy, rendering that variant more stable. The contour line is observed to be wiggling, confirming energetically an interfusing occupation of the orientation variants in phase space originating from the high rotational freedom of BA molecules.

In general, variant I will be stabilized under compression (tension) along $a$ ($b$), whereas tension

(compression) along *a* (*b*) would promote the stability of variant II. Notably, in the regime of pure compression (see lower-left region of Fig. 2(b)), variant II can be restabilized with increasing compression along *a*. This may be linked to the dominant role of the out-of-plane octahedral tilting mode during compression in minimizing the strain energy. By analyzing the buckling height of the equatorial I atoms $\delta_{eq}$, which is taken as a measure of out-of-plane tilting (Fig. S3(a)), the anisotropic response of the perovskite structure in this regime is uncovered (see lower-left regions of both panels in Fig. S3(b)). As shown, out-of-plane octahedral tilting will be initiated in variant I (II) with increasing compression along *b* (*a*), but not along *a* (*b*). Since the out-of-plane tilting mode will be activated in variant II but not in variant I with increasing compression along *a*, the strain energy could be effectively lowered in variant II to make it the more stable variant locally.

The energy surfaces of the two variants can be seen to intersect within 1–2% of axial strain, which suggests that ferroelastic switching can be experimentally achievable by applying low to moderate stresses in the range of ~0.12–0.24 GPa (*26*). Furthermore, like other 2D layered HRPPs, monolayer (BA)$_2$PbI$_4$ is mechanically soft and flexible in nature, which makes it highly responsive to mechanical loading and implies reduced energy barriers of structural changes (*26, 27*). Stress (and consequently, strain) could be applied to the monolayer *via* the deformation of a flexible substrate, as exemplified in TMDs and other HRPPs (*17, 28*). Another approach, which might yield more quantitative evidence of the switching process, is to perform an atomic force microscopy experiment where a probe pushes down onto a freely suspended monolayer across a trench (*29, 30*). A ferroelastic transition, from variant I to II for instance, would however involve significant rotations of BA molecule pairs (see Fig. 2(a)). Hence, whether or not orientation variant switching can ultimately take place would depend on the barriers to rotation of the BA molecules, which in turn would hinge on their rotational coupling mode across the Pb–I inorganic framework. In the next subsection, we will systematically explore the potential FE transition pathways in monolayer (BA)$_2$PbI$_4$, including the energy barriers to switching.

**Modes of Molecular Rotation and Mechanism of Ferroelastic Switching**

Ferroelastic switching in conventional 2D-FE materials such as phosphorene, TMDs, and group-IV monochalcogenides have been attributed to localized, displacive movements of a group of atoms. However, the mechanism of ferroelastic switching in 2D layered HRPPs is different as it is mainly driven by the rotation of interior-site cations, for instance, MA molecules. Likewise, in the present study on (BA)$_2$PbI$_4$, ferroelasticity is found to be primarily driven through rotations of the BA organic spacer molecules. Although our above results may suggest that it becomes energetically favorable for (BA)$_2$PbI$_4$ monolayers to switch between the I and II variants under small strains, if the kinetic barrier associated with variant switching is too high, such variant switching may not occur under normal experimental conditions and timescale, and the material would still not exhibit practical ferroelastic qualities.

In determining the kinetic barriers of ferroelastic switching, we must first consider the potential pathways by which the BA molecule pairs can undergo rotation. Each BA molecule comprising the pair has two degrees of rotational freedom, where it can either rotate in a clockwise manner or an anticlockwise manner. Based on these degrees of freedom, three fundamental, symmetry-inequivalent coupling modes can be identified, referred to here as the positive in-phase, negative in-phase, and out-of-phase rotational mode (defined relative to variant I). The schematics of molecular rotation for each mode (on a per unit cell basis) are as shown in the right panels of Figs. 3(a)–(c). In the positive in-phase rotational mode, both molecules rotate anticlockwise, whereas in the negative in-phase rotational mode, both molecules rotate clockwise. As for the out-of-phase rotational mode, one molecule rotates anticlockwise, while the other rotates clockwise.

Ferroelastic switching under each of these three modes are computed using the climbing image nudged elastic band (CINEB) (*31*) method to investigate the transformation pathway and kinetic barriers to switching. Two independent calculations are performed, one based on the groundstate lattice constants of variant I and the other on the groundstate lattice constants of variant II. Subsequently, a fitted curve to the image data points is obtained to evaluate the kinetic barriers, as shown in the left panels of Figs. 3(a)–(c).

Interestingly, the saddle points in the CINEB results reveal the existence of metastable states for each of the transformation pathways corresponding to distinct paraelastic phases of monolayer $(BA)_2PbI_4$, which we denote as $IP^+$, $IP^-$, and OOP. This is unlike conventional ferroelasticity whereby the PE phase is usually unstable. The energy profiles obtained based on the groundstate lattice constants of variants I (in red) and II (in blue) closely overlap, indicating that the predicted barriers are insensitive to the lattice parameters of the simulation cell and are representative of the transition. From the fitted curve, the kinetic barriers to FE switching can be determined using two metrics, the activation barrier $E_b^{I/II}$ and the decomposition barrier $E_d^{I/II}$, as defined in Fig. 3(d) (upper panel). The predicted energy barriers for each rotational mode are plotted in the lower panel of Fig. 3(d). The $E_b^{I/II}$ values are in the range of 0.46–0.76 eV, with the lowest activation barrier corresponding to the out-of-phase rotational mode, implying that it could offer the most facile ferroelastic switching pathway. The OOP metastable state is also found to be the least stable, with a relatively low $E_d^{I/II}$ of around 0.13 eV, while the in-phase metastable states $IP^+$ and $IP^-$ are in comparison highly stable, with $E_d^{I/II}$ values of around 0.31–0.46 eV. The higher $E_d^{I/II}$ values of the $IP^+$ and $IP^-$ metastable states suggest their longer lifetimes, and thus, these PE phases could even be observed experimentally and will principally form paraelastic–ferroelastic domain boundaries. At the same time, it is worth noting that the barriers to molecular rotation can be expected to be higher for few-layer $(BA)_2PbI_4$ due to the additional energy required to overcome the interlayer vdW forces.

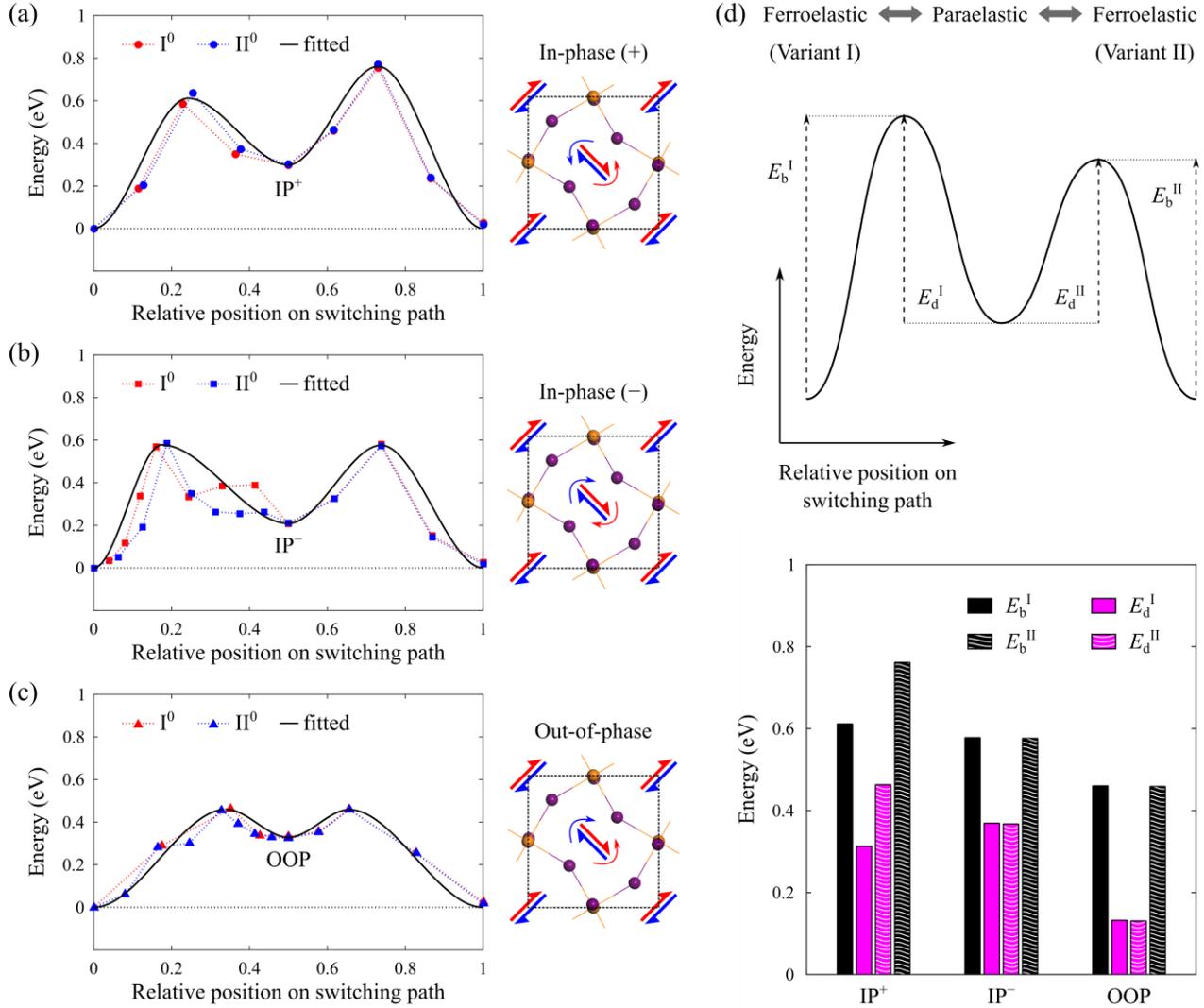

**Fig. 3.** (a)–(c) CINEB profiles (left panel) and top-view structural schematics (right panel) of ferroelastic switching from variant I to II for the positive in-phase, negative in-phase, and out-of-phase rotational mode, respectively. (d) Definition of the activation and decomposition barriers associated with ferroelastic switching (upper panel) and their values for each rotational mode (lower panel).

**Structural and Electronic Effects of Metastable Reordering**

Our calculations unveil the existence of different metastable PE states generated during FE switching depending on the rotational coupling mode of BA pairs across the Pb–I inorganic lattice. Particularly, the structural and electronic effects of metastable reordering is worthy of further investigation because such PE states in $(BA)_2PbI_4$ can have long lifetimes and contribute to the perovskite's multidomain structure. In that sense, proper control over the concentration and distribution of metastable states may also offer an exciting avenue for tuning the functional properties of the material in practice.

The fully relaxed configurations of the various PE phases: $IP^+$, $IP^-$, and OOP, are obtained, as illustrated in Figs. 4(a)–(c). The $IP^+$ phase comprises BA molecules aligned entirely in a head-to-tail manner along both $x$ and $y$, while, in the $IP^-$ phase, the BA molecules are oriented purely in a head-to-head manner along the principal axes. As for the OOP phase, along both $x$ and $y$, the BA molecules are aligned in a head-to-tail manner across one of the surfaces but in a head-to-head manner across the other. This gives rise to a net polarization along the [1 1 0] direction, which may explain why the OOP metastable state is relatively unstable compared to the in-phase metastable states $IP^+$ and $IP^-$, in which polarization is quenched. In view of the isotropic arrangement of atoms in all the identified PE phases, $IP^+$, $IP^-$, and OOP can all be regarded as tetragonal structures; their optimized lattice constants ($a = b$) are evaluated to be 8.50 Å, 8.61 Å, and 8.60 Å, respectively. Since the head-to-head alignment of BA molecules is generally responsible for a dimensionally expanded structure due to steric effects, its lack thereof in the $IP^+$ phase can be seen to lead to appreciably smaller lattice dimensions.

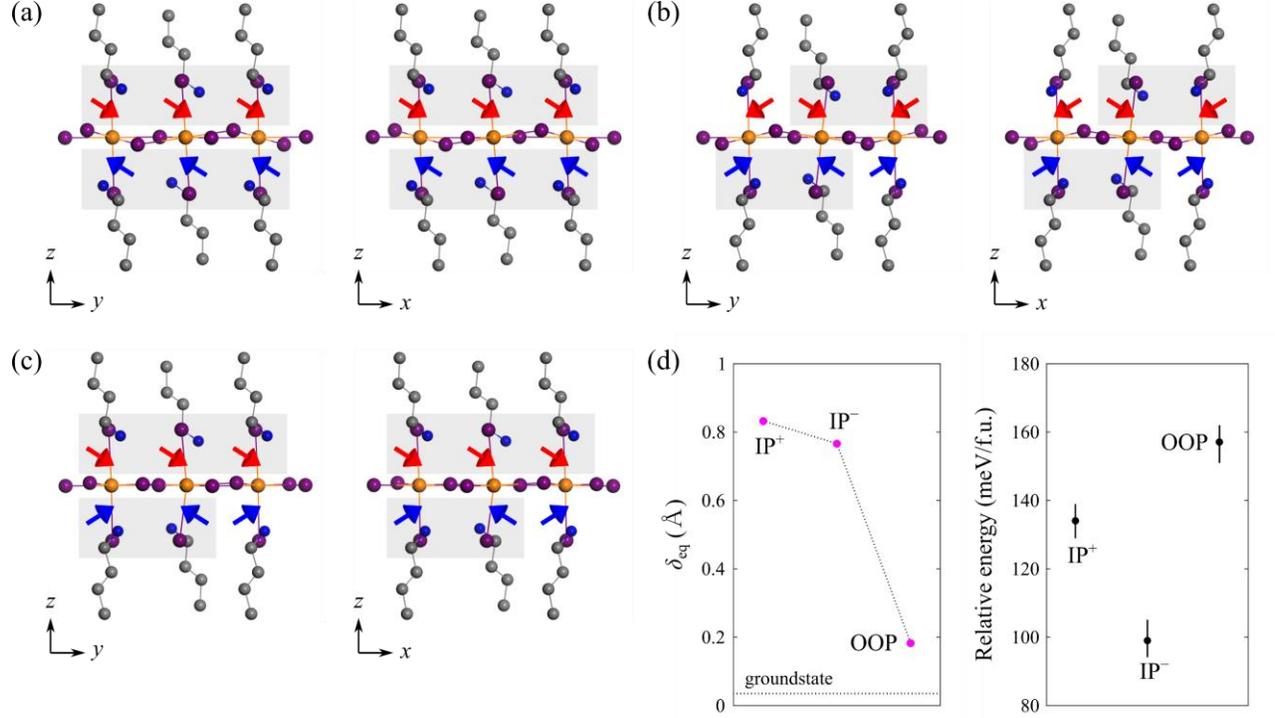

**Fig. 4.** Optimized crystal structures of the paraelastic phases: (a) IP$^+$, (b) IP$^-$, and (c) OOP. (d) The buckling height $\delta_{eq}$ (left panel) and relative energy (right panel) associated with each paraelastic phase.

The spontaneous lattice strain in each of the ferroelastic orientation variants I and II can be mathematically described by a 2 × 2 in-plane transformation strain matrix with the PE phase taken as the reference state. To this end, a paraelastic-to-ferroelastic transformation strain matrix $\boldsymbol{\eta}$ can be defined using the Green–Lagrange strain tensor formalism as

$$\boldsymbol{\eta} = \frac{1}{2}\left[\left(\mathbf{H}_{PE}^{-1}\right)^T \mathbf{H}_{FE}^T \mathbf{H}_{FE} \mathbf{H}_{PE}^{-1} - \mathbf{I}\right], \qquad (2)$$

where $\mathbf{H}_{PE}$ and $\mathbf{H}_{FE}$ denote the unit cell matrices of the paraelastic and ferroelastic phases, respectively, and $\mathbf{I}$ is a 2 × 2 identity matrix. The transformation strain matrices describing the spontaneous structural evolution of monolayer (BA)$_2$PbI$_4$ from its PE phase (IP$^+$, IP$^-$, or OOP) to either variant I or II in its FE phase can thus be derived as follows:

$$\boldsymbol{\eta}_{\mathrm{IP}^+}^{\mathrm{I}} = \begin{bmatrix} 0.0166 & 0 \\ 0 & 0.0236 \end{bmatrix}, \quad \boldsymbol{\eta}_{\mathrm{IP}^+}^{\mathrm{II}} = \begin{bmatrix} 0.0236 & 0 \\ 0 & 0.0166 \end{bmatrix}, \quad (3.1)$$

$$\boldsymbol{\eta}_{\mathrm{IP}^-}^{\mathrm{I}} = \begin{bmatrix} 0.0031 & 0 \\ 0 & 0.0099 \end{bmatrix}, \quad \boldsymbol{\eta}_{\mathrm{IP}^-}^{\mathrm{II}} = \begin{bmatrix} 0.0099 & 0 \\ 0 & 0.0031 \end{bmatrix}, \quad (3.2)$$

$$\boldsymbol{\eta}_{\mathrm{OOP}}^{\mathrm{I}} = \begin{bmatrix} 0.0044 & 0 \\ 0 & 0.0112 \end{bmatrix}, \quad \boldsymbol{\eta}_{\mathrm{OOP}}^{\mathrm{II}} = \begin{bmatrix} 0.0112 & 0 \\ 0 & 0.0044 \end{bmatrix}, \quad (3.3)$$

where the subscript (superscript) following $\boldsymbol{\eta}$ refers to the specific PE (FE) configuration of interest.

The reorientation of BA molecules in the metastable PE phases induces perceptible distortions in the Pb–I octahedra network, with stronger out-of-plane tilting of the octahedra in the $\mathrm{IP}^+$ and $\mathrm{IP}^-$ phases than in the OOP phase, as reflected by the extent of buckling of the equatorial I atoms $\delta_{\mathrm{eq}}$ (Fig. 4(d), left panel). There exists significant buckling of about 0.8 Å in the $\mathrm{IP}^+$ and $\mathrm{IP}^-$ configurations, while that of the OOP structure is only around 0.2 Å, close to that of the FE groundstate. To assess the relative stability of each of the PE phases, we calculate its relative energy (Fig. 4(d), right panel), which is defined as the energy difference between the fully relaxed PE configuration and the FE groundstate. The $\mathrm{IP}^-$ configuration appears to be the most stable PE phase, with the lowest relative energy of ~99 meV/f.u., whereas that of the OOP structure is the highest (~157 meV/f.u.), and it could easily transit to the FE phase.

Bearing in mind that the valence and conduction bands of the perovskite are dominated by states contributed by atoms in the inorganic layer (Fig. S2), the structural changes associated with metastable reordering can be expected to trigger strong alterations in the electronic band alignment, as shown in Fig. 5. The band gap and work function of monolayer $(\mathrm{BA})_2\mathrm{PbI}_4$ in its FE groundstate is evaluated to be 2.12 eV and 4.84 eV, respectively. Metastable PE reordering results in a general upward shift of the energy levels, leading to smaller work functions of around 4.63–4.69 eV. As depicted in Fig. 5, the band gap is slightly blue-shifted in the IP$^+$ and OOP phase, though it retains a similar value of 2.13 eV in the IP$^-$ phase. Importantly, by driving higher energy states, metastable reordering could induce a type-II band alignment formed between the PE and FE phase. This implies that metastable PE domains can act as sinks for funneling photo-generated holes, reminiscent of exciton funnels formed in rippled 2D layered black phosphorus (*32*). In contrast, photo-generated electrons will tend to move towards the FE domains of the perovskite, thus promoting electron–hole separation across PE–FE domain boundaries.

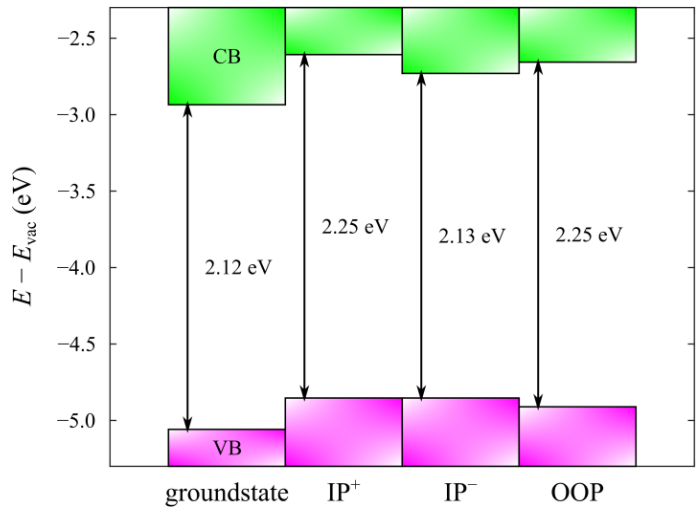

**Fig. 5.** Valence (pink) and conduction (green) bands, abbreviated as VB and CB, respectively, of the ferroelastic groundstate and paraelastic phases, aligned relative to the vacuum level $E_{vac}$.

**Effect of Thermal Excitation and Halogen Mixing on the Activation of Ferroelastic Switching**

To ascertain the potential impact of thermal excitation on the activation of ferroelastic switching, we conducted metadynamics simulations of monolayer (BA)$_2$PbI$_4$ at 100, 200, and 300

K. The simulations were carried out using two types of collective variables (CVs): (i) the lattice constants ($a$ and $b$), and (ii) the angle formed between two BA molecules comprising a cross-plane pair, as defined in Fig. 6(a). Our investigation reveals that thermal effects are pivotal in governing the FE switching process. Our analysis, as presented in Fig. 6(b), involved tracking the lattice variation along both $a$ and $b$ directions, and we observe that an increase in temperature results in a lower energy requirement for lattice dilation. This is indicative of increased molecular mobility at higher temperatures which can facilitate thermally induced rotations of the BA molecules. A closer inspection *via* distribution analysis, as shown in Figs. 6(c)–(e), confirms the possibility of lattice expansion and contraction due to thermal fluctuations and suggests that FE switching will be favored at elevated temperatures.

To determine the exact temperature-dependent $E_b$ of FE switching, we consider the relative spatial alignment of two cross-plane coupled BA molecules and employ the angle formed between their linear backbones as the CV (see Fig. 6(a)). The resulting free energy plots, as depicted in Figs. 6(g)–(i), exhibit a consistent declining trend whereby the rotation of the two BA molecules is more unimpeded at elevated temperatures, specifically at 300 K, where the energy barrier is only 17.0 kJ/mol. In contrast, at lower temperatures, $E_b$ increases, reaching 27.5 kJ/mol at 200 K and 39.4 kJ/mol at 100 K. Therefore, temperature plays an important role in governing ferroelastic switching.

To investigate the potential impact of halogen mixing, which has been proven to be an effective means to enhance the properties of HRPPs, we randomly substituted half of the I atoms with Br atoms (*i.e.*, 50% halide replacement). Bromine substitution is revealed to have a strong effect on the lattice constants and rotational freedom of the organic molecules. Evidently, for the case of 50% halide replacement at 300 K, the lattice constants are constrained below ~12.5 Å along both *a* and *b* directions (Fig. 6(f)). This upper bound is smaller than that of the pristine case with purely I atoms, where the lattice constants can reach up to 13.0 Å (Fig. 6(e)). The reduced lattice constants impart a steric effect on the BA molecules, resulting in an increase in the $E_b$ from 17.0 kJ/mol to 35.7 kJ/mol (Figs. 6(i) and 6(j)). As ferroelastic switching relies heavily on the rotational freedom of the organic molecules, this elevation in the barrier implies a constraint on the switching dynamics compared to the scenario with purely I atoms.

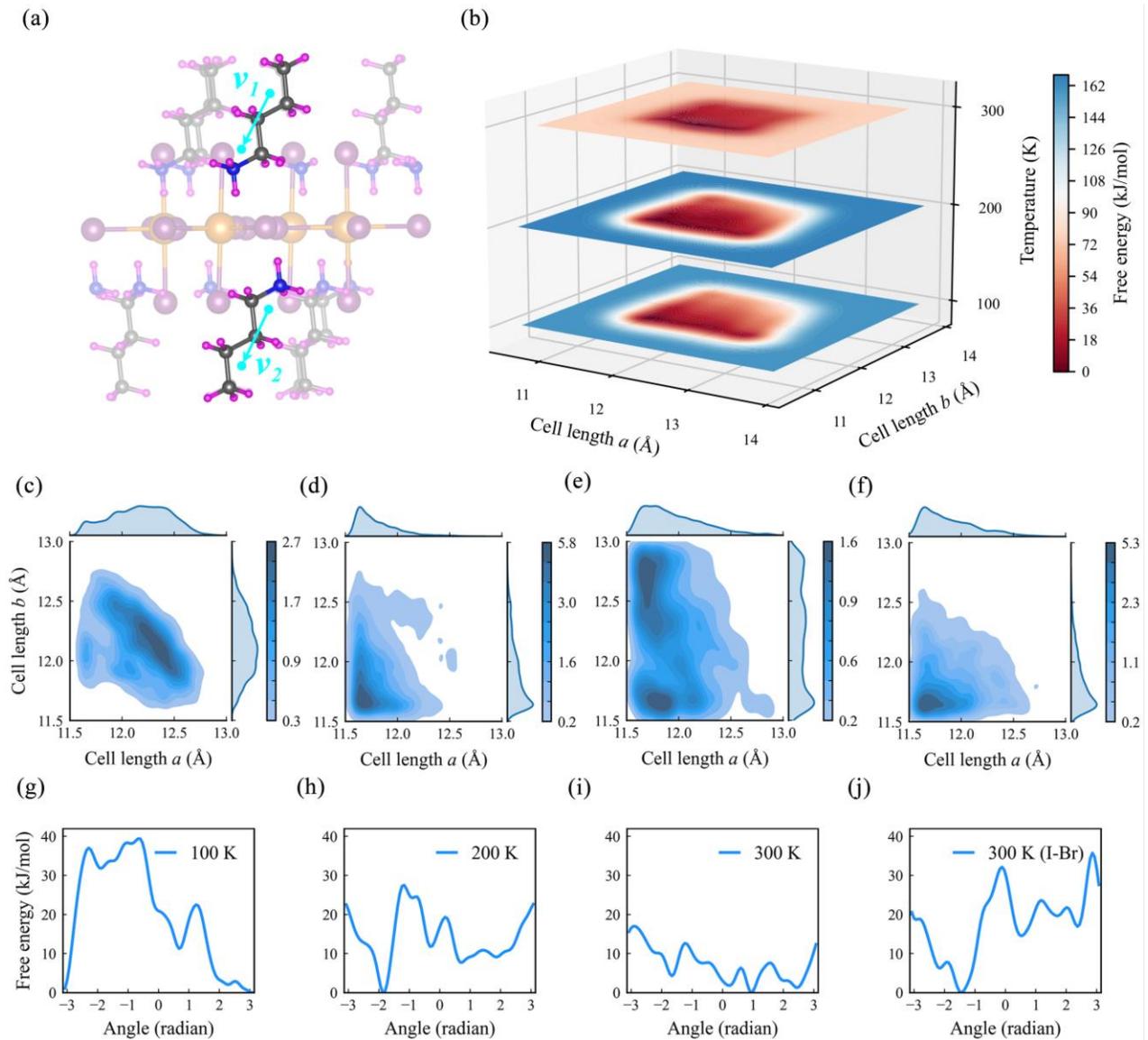

**Fig. 6.** Effect of thermal excitation and halogen mixing on the activation of ferroelastic switching.

(a) Structural configuration of monolayer $(BA)_2PbI_4$, where the vectors $v_1$ and $v_2$ (cyan) define the orientation of a BA molecule located on the top and bottom surface of the monolayer, respectively. Note that H atoms are indicated in pink. (b) Free energy surface of monolayer $(BA)_2PbI_4$ at different temperatures, obtained *via* metadynamics simulations with the lattice constants taken as the CV. (c)–(f) Lattice variation distribution at 100 K, 200 K, 300 K, and 300 K with 50% halide replacement where I atoms are randomly substituted by Br atoms, respectively. The color bar exhibits the kernel density of distribution. (g)–(j) Free energy plot of molecular coupling at 100 K, 200 K, 300 K, and 300 K with 50% halide replacement where I atoms are randomly substituted by Br atoms, respectively, obtained *via* metadynamics simulations with the angle formed between two cross-plane coupled BA molecules taken as the CV.

Given that the organic BA molecules are strongly coupled to the inorganic lattice, the molecular orientation-induced FE transition is always accompanied by octahedral tilting. With reference to previous literature (*33*), azimuthal angles ($\sigma$, $\varphi$) are introduced here to describe the structural transformation of the FE variants and PE phases, as shown in Fig. S4(a). The energy barriers for octahedral tilting with respect to a certain direction (*e.g.*, the *z*-axis, which corresponds to the distribution density of ($\sigma_z$, $\varphi_z$)) can be extracted from the *ab initio* molecular dynamics (AIMD) trajectories of the various structures, obtained at 200 K, 300 K, and 400 K. As shown in Fig. S5, the anharmonic motion of heavy lead atoms causes the energy surface to deviate from the harmonic potential, leading to double energy peaks for both the in-plane (*x* and *y*) and out-of-plane (*z*) directions rather than a unimodal peak. The FE variants I and II have relatively low energy barriers for tilting compared to the metastable IP$^+$ phase or OOP phase as the dynamic tilting process can be hindered in the latter two systems with smaller lattice constants. Furthermore, such tilting shows a clear temperature-dependent feature, whereby the barrier height of the azimuthal angles along *x* of variant I decreases from 0.25 Kcal/mol at 200 K to 0.16 Kcal/mol at 400 K (left and center panels of Fig. S5). Due to the strong lattice anharmonicity, local energy minima are formed at low temperatures and the joint distributions reveal double distribution centers of the tilting angles at 200 K (right panel of Fig. S5). Increased temperatures meanwhile lead to the system gaining energy to overcome the tilting barrier, with enhanced atomic motion contributing to a wider range and stronger unimodality of the angle distribution at 300 and 400 K. Similar plots of the temperature-dependent barrier heights and angle distribution are provided in Figs. S6–S8 for the other FE variant and PE phases. According to the phonon density of states distribution of the monolayer (Fig. S9), we can see that there are two primary peaks at ~1450 $cm^{-1}$ and ~3000 $cm^{-1}$ related to the organic–inorganic coupled phonon modes. The former corresponds to the torsion of the carbon bone and bending of the terminal $NH_3^+$ group, while the latter is linked to the stretching of the $NH_3^+$ group towards the inorganic Pb–I lattice. In particular, the peak at ~3000 $cm^{-1}$ exhibits a temperature correlation that favors enhanced stretching of the $NH_3^+$ group at higher temperatures. This may prompt stronger interactions with the inorganic lattice, hence lowering the barrier heights.

Organic cation reorientation-related entropy changes are the primary contributor to the order–disorder type phase transition in bulk and monolayer perovskites (*34–36*). Using an auxiliary axis component of the C–N vector ($\omega$ vector; see Figs. S4(b)–(f)) to indicate the molecular orientation within the dynamic process, the temperature-dependent behavior of the organic part is revealed. The FE variants I and II have almost equivalent $x$ and $y$ components of the $\omega$ vectors at 0 ps, and thermal vibrations only lead to small deviations, which implies robust ferroelastic phase stability of these variants in the absence of strain (Figs. S10 and S11). Meanwhile, heating the same systems above 300 K leads to the fast swapping of axis components and the initial direction characters are replaced by an isotropic distribution stage, same as that observed for the intermediate $IP^+$ and OOP phases. For comparison, the $\omega$ vectors of the intermediate $IP^+$ and OOP phases have merely an $x$ component (Fig. S12) initially at 0 ps. Atomic thermal fluctuations, however, cause the reorientation of the BA cations, and the final *y* component evolves over time to become close to the *x* component, indicating an isotropic distribution tendency.

**Conclusions**

This article reports on the existence of dual-state ferroelasticity in $(BA)_2PbI_4$, an archetypical member of the $n = 1$ HRPP family. Based on a comprehensive theoretical study of the monolayer, we reveal that ferroelastic switching is mediated by cross-plane intermolecular coupling, which could occur through multiple rotational modes, rendering the formation of FE domains and several metastable paraelastic phases. By analyzing the energy landscape of the FE phase, it is shown that the application of strain, by as little as a few percent, can promote orientation variant switching. In addition, as a general trend, molecular rotations associated with FE switching appear to be facilitated with increasing temperature. This finding is corroborated by a number of previous experiments on hybrid perovskites which have reported on thermally activated perturbations of the organic spacer molecules (*18–22*). Notably, metastable reordering could induce a type-II band alignment formed between the PE and FE phase. Such an alignment would favor the splitting of electrons and holes, and in this way, PE states can play a similar role as that of polarization in ferroelectrics in terms of boosting charge carrier splitting. The coexistence of PE and FE domains can therefore promote electron–hole separation across their boundaries, potentially enhancing the photovoltaic performance of the perovskite. Besides, tensile edge stresses have been reported in monolayer $(BA)_2PbI_4$ (*26*), which suggest that PE phases with smaller lattice dimensions can be stabilized near the edge rather than in the interior of the nanosheet. In a recent work by Xiao *et al.*, it was observed experimentally that $(BA)_2PbI_4$ crystals with thicknesses of ~10 μm were not ferroelastic (*17*). While this was presumed to be due to the absence of interior-site cations, on the basis of our investigation, it could be instead argued that ferroelasticity may vanish in such thick-film (or bulk) structures because the BA molecules are locked in place by interlayer vdW forces, restricting rotational modes which should normally contribute to ferroelasticity, as seen in the case of the monolayer. The present findings affirm that the origins of ferroelasticity in $n = 1$ HRPPs vastly differ from that in other 2D-FE materials involving displacive atomic motion, and to some extent, even from $n > 1$ HRPP systems, where ferroelasticity is instead initiated by the rotational coupling of small interior-site cations.

Halogen mixing is demonstrated as a viable means for regulating the activation of ferroelastic switching in $(BA)_2PbI_4$. In the context of our results, Br substitution tends to lead to reduced lattice constants, which impart a steric effect on the BA molecules, resulting in an increase in the rotation barrier. Though changes in the hydrogen bonding strength have not been explicitly considered in our study, they are expected to be just as likely to affect the barriers to rotation and would offer an interesting perspective as part of future work. Another plausible channel for realizing FE tunability would be to replace some of the short-chain BA molecules with an alternative organic species. For example, longer-chain aliphatic molecules are known for having much more flexible characters (*37*, *38*), which could be helpful for facilitating molecular reorientation. Moreover, compared to primary amines, secondary amines might have different, and perhaps, even more docking possibilities, hence modifying the energy landscape enabling ferroelastic response. The changes in hydrogen bonding of secondary amines could also alter the strength of intermolecular coupling and may endow the molecules with greater rotational freedom. Compositional engineering through halogen and/or organic molecule mixing could thus offer rich avenues for tuning the ferroelastic performance of HRPPs. It is worth noting that the functional properties of the perovskite are ultimately structure-dependent and will vary in accordance with the different variants and phases stable in the system, which are in turn subject to their energetics. Ideally, if the FE phase can be reversibly transformed to the PE phase under external stimuli, be it under thermal or mechanical fields, (BA)$_2$PbI$_4$ monolayers could then serve as molecularly thin shape memory materials, with operating principles similar to traditional shape memory alloys (*39*). Furthermore, the domain boundaries formed between FE variants or between PE and FE phases, which are quasi-one-dimensional defects in 2D materials, may also possess exotic physics and provide fertile ground for domain boundary engineering (*1*, *4*). Overall, our study makes significant headway in the quest of harnessing ferroelastic phenomena in ultrathin HRPPs, and the predicted results provide strong motivation as well as key theoretical guidance for chemists and physicists working in this exciting field to further unveil the intricacies of organic molecule reordering and organic–inorganic lattice coupling.

**Methods**

**First-Principles Calculations**

Spin-polarized first-principles calculations are performed based on the Perdew–Burke–Ernzerhof (PBE) (*40*) exchange–correlation functional within the framework of DFT (*41*, *42*) using the Vienna *ab initio* simulation package (VASP) (*43*). The DFT-D2 method of Grimme (*44*) is applied to describe the vdW interactions in 2D layered (BA)$_2$PbI$_4$. A kinetic energy cutoff of 500 eV is selected for the plane wave basis set. Periodic boundary conditions are applied in the in-plane ($x$, $y$) directions, whereas free boundary conditions are imposed in the normal ($z$) direction by introducing a vacuum separation layer of ~15 Å between neighboring slabs. The investigated monolayers are sampled in the Brillouin zone with a $4 \times 4 \times 1$ *k*-point grid using the Monkhorst–Pack method. The energy convergence criteria for electronic iterations is set at $10^{-6}$ eV, and all structures are relaxed until the maximum Hellmann–Feynman force per atom is smaller than 0.02 eV/Å. The CINEB (*31*) method is adopted to investigate the ferroelastic switching pathways and barriers to molecular rotation in the system.

**Molecular Dynamics Simulations**

To obtain the free energy, AIMD and *ab initio* metadynamics (AIMTD) simulations are implemented through the i-PI (*45*) software with the PLUMED (*46*) plugin, and VASP (*43*) is used as the force engine. A $\sqrt{2} \times \sqrt{2} \times 1$ supercell is adopted for all AIMD and AIMTD simulations. Only the Gamma point is used for sampling the Brillouin zone. All AIMD simulations are performed with a timestep of 0.25 fs. As for the AIMTD computational routine, a 10.0 ps-long AIMD simulation is first performed under an isothermal–isochoric (NVT) ensemble to fully equilibrate the system, followed by an AIMTD simulation under an isothermal–isobaric (NPT) ensemble with Langevin dynamics (time constant: 0.1 ps) and a timestep of 1.0 fs to explore the free energy surface. The AIMTD simulation employs Gaussian "hills" with a deposition rate of 20 fs, width of 0.25 Å, and height of 1.0 kJ/mol. Two types of CVs are selected: (i) the lattice constants (*a* and *b*) of the supercell, and (ii) the angle formed between two BA molecules comprising a cross-plane pair. For each type of CV, we conducted AIMTD simulations at temperatures of 100 K, 200 K, and 300 K. Here, we represent the orientation of the BA molecule located on the top (bottom) surface of the monolayer by a vector $v_1$ ($v_2$), as follows:

$$v_1 = C_2 - C_1, \qquad (4.1)$$

$$v_2 = C_1 - C_2, \qquad (4.2)$$

where $C_1$ and $C_2$ are two centers of mass corresponding to the C–C–C and C–C–N segments that form the backbone of the organic molecule, respectively.

We then calculate the projected angle between the vectors $v_1$ and $v_2$ on the $a$–$b$ plane ($\theta$), as follows:

$$\theta = \cos^{-1}\left(\frac{v_1^* * v_2^*}{|v_1^*||v_2^*|}\right), \quad (5)$$

where $v_1^*$ and $v_2^*$ are the projected vectors of the BA molecules on the $a$–$b$ plane.

Besides, the molecular reorientation-related energy is determined based on the probability distribution of azimuthal angles $\rho(\sigma, \varphi)$, and can be expressed as (*33*)

$$F(\sigma_i) = -k_B T \ln\left(\int d\varphi_i \rho(\sigma_i, \varphi_i)\right), \quad (6.1)$$

$$F(\varphi_i) = -k_B T \ln\left(\int d\sigma_i \rho(\sigma_i, \varphi_i)\right), \quad (6.2)$$

where $\sigma_i$ and $\varphi_i$ denote the azimuthal angles with respect to a certain direction ($i = x$, $y$, or $z$).

**Supporting Information**

The Supporting Information is available free of charge at [*insert link here*]. The file contains additional first-principles simulation results, schematics of the crystal structure of monolayer (BA)$_2$PbI$_4$, and molecular dynamics simulation results accessory to the main article (PDF).

**Conflict of Interest Disclosure**

The authors declare no competing financial interest.

**Acknowledgments**

K. Zhou acknowledges the financial support received from the Nanyang Environment and

Water Research Institute (Core Funding), Nanyang Technological University, Singapore. Y. Cai acknowledges the support provided by the Science and Technology Development Fund from Macau SAR (0120/2023/RIA2, 0085/2023/ITP2), the Natural Science Foundation of China (grant 22022309), and the Natural Science Foundation of Guangdong Province, China (2021A1515010024). The computational work for this article was partially performed on resources of the National Supercomputing Centre, Singapore (https://www.nscc.sg).

Method for Finding Saddle Points and Minimum Energy Paths. *The Journal of Chemical Physics* **2000**, *113* (22), 9901-9904.

32. Quereda, J.; San-Jose, P.; Parente, V.; Vaquero-Garzon, L.; Molina-Mendoza, A. J.; Agraït, N.; Rubio-Bollinger, G.; Guinea, F.; Roldán, R.; Castellanos-Gomez, A., Strong Modulation of Optical Properties in Black Phosphorus through Strain-Engineered Rippling. *Nano Letters* **2016**, *16* (5), 2931-2937.

33. Kaiser, W.; Carignano, M.; Alothman, A. A.; Mosconi, E.; Kachmar, A.; Goddard, W. A., III; De Angelis, F., First-Principles Molecular Dynamics in Metal-Halide Perovskites: Contrasting Generalized Gradient Approximation and Hybrid Functionals. *The Journal of Physical Chemistry Letters* **2021**, *12* (49), 11886-11893.

34. Deng, W.-F.; Li, Y.-X.; Zhao, Y.-X.; Hu, J.-S.; Yao, Z.-S.; Tao, J., Inversion of Molecular Chirality Associated with Ferroelectric Switching in a High-Temperature Two-Dimensional Perovskite Ferroelectric. *Journal of the American Chemical Society* **2023**, *145* (9), 5545-5552.

35. Leng, K.; Li, R.; Lau, S. P.; Loh, K. P., Ferroelectricity and Rashba Effect in 2D Organic–Inorganic Hybrid Perovskites. *Trends in Chemistry* **2021**, *3* (9), 716-732.

36. Shi, P.-P.; Lu, S.-Q.; Song, X.-J.; Chen, X.-G.; Liao, W.-Q.; Li, P.-F.; Tang, Y.-Y.; Xiong, R.-G., Two-Dimensional Organic–Inorganic Perovskite Ferroelectric Semiconductors with Fluorinated Aromatic Spacers. *Journal of the American Chemical Society* **2019**, *141* (45), 18334-18340.

37. Dhanabalan, B.; Biffi, G.; Moliterni, A.; Olieric, V.; Giannini, C.; Saleh, G.; Ponet, L.; Prato, M.; Imran, M.; Manna, L.; Krahne, R.; Artyukhin, S.; Arciniegas, M. P., Engineering the Optical Emission and Robustness of Metal-Halide Layered Perovskites through Ligand Accommodation. *Advanced Materials* **2021**, *33* (13), 2008004.

38. Tu, Q.; Spanopoulos, I.; Hao, S.; Wolverton, C.; Kanatzidis, M. G.; Shekhawat, G. S.; Dravid, V. P., Out-of-Plane Mechanical Properties of 2D Hybrid Organic–Inorganic Perovskites by Nanoindentation. *ACS Applied Materials & Interfaces* **2018**, *10* (26), 22167-22173.

39. Bhattacharya, K., *Microstructure of Martensite: Why it Forms and How it Gives Rise to the Shape-Memory Effect*. Oxford University Press, 2003; Vol. 2.

40. Perdew, J. P.; Burke, K.; Ernzerhof, M., Generalized Gradient Approximation Made Simple.

**For Table of Contents Only**

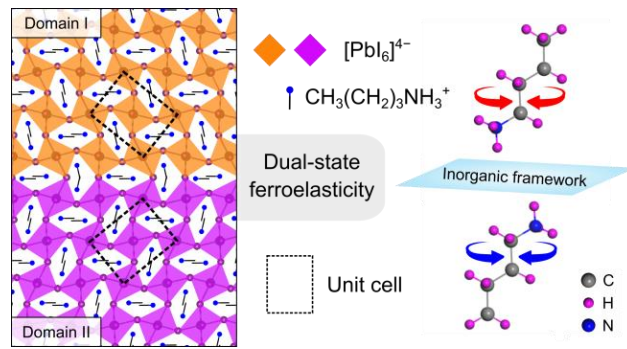